\documentclass[12pt]{article}

\usepackage{graphicx}
\usepackage{latexsym}

\begin{document}

\newcommand{\preprintno}[1]
{{\normalsize\begin{flushright}#1\end{flushright}}}

\title{\preprintno{BA-TH/00-402 \\ gr-qc/0101019}
       \quad \\
       Gravitational waves in non-singular string cosmologies}

\author{C.~Cartier\thanks{E-mail: c.cartier@sussex.ac.uk}, \,
        E.J.~Copeland\thanks{E-mail: e.j.copeland@sussex.ac.uk}\\
        {\em Centre for Theoretical Physics, University of Sussex,} \\
        {\em Falmer, Brighton BN1 9QJ, U.~K.} \\
        \quad \\
        and \\
        \quad \\
        M.~Gasperini\thanks{E-mail: gasperini@ba.infn.it}\\
        {\em Dipartimento di Fisica, Universit\`{a} di Bari,} \\
        {\em Via G. Amendola 173, 70126 Bari, Italy.} \\
        \quad}

\date{5th January 2001}

\maketitle
\begin{abstract}
We study the evolution of tensor metric fluctuations in a class of
non-singular models based on the string effective action, by including
in the perturbation equation the higher-derivative and loop corrections
needed to regularise the background solutions. We discuss the effects
of such higher-order corrections on the final graviton spectrum, and we
compare the results of analytical and numerical computations.
\end{abstract}

\newpage

\section{Introduction}
Recent studies, including strong coupling corrections to the
string effective action, have provided the pre-Big Bang scenario
\cite{Veneziano:1991ek,Gasperini:1993em} with a number of promising models for the
transition from the growing to the decreasing curvature regime
\cite{Brustein:1998cv,Foffa:1999dv,Brustein:1999yq,Cartier:1999vk}.
However, a crucial question remains: is it possible to test this
cosmological scenario and, in particular, to distinguish it from
other, more conventional descriptions of the very early Universe?

The answer to this essential question is in principle positive, since the
pre-Big Bang scenario incorporates a dilaton field
which modifies the standard inflationary kinematics and which,
through its non-minimal coupling, also directly modifies
the evolution of  perturbations
(of the metric, and of the other background fields) \cite{Gasperini:1994xg}.
It is well known, on the other hand, that the transition from an
inflationary period to the usual Friedman-Robertson-Walker (FRW)
phase amplifies tensor metric perturbations, and is associated with
the production of a stochastic background of relic gravitational
waves \cite{Grishchuk:1975ny,Starobinsky:1979ty}.

Such a primordial background decouples from matter
immediately below the Planck scale,
unlike the electromagnetic radiation which underwent a
complicated history until recombination, and has been transmitted
almost unperturbed down to our present epoch. As a
consequence, its present spectrum should be a faithful portrait of
the very early universe \cite{Grishchuk:1991kk}, thus opening a window for the
observation of processes occurring near to the Planck scale, and for
discriminating among various models of primordial evolution.

In the context of string cosmology,
several mechanisms may contribute to the generation of gravitational
waves from the initial vacuum state.
One contribution comes from the process of dynamical dimensional
reduction \cite{Garriga:1989jx,Demianski:1990ay,Gasperini:1992sm}, when the internal
dimensions shrink down to a final compactification scale. Another
contribution is due to the time-dependence of the dilaton field --
and thus of the effective gravitational coupling constant --
during the pre-Big Bang phase \cite{Gasperini:1993dp,Barrow:1993ad}.
Finally, there is the usual contribution arising from the accelerated
expansion of the external three-dimensional space. Both the dilaton
and the full, higher-dimensional metric thus contribute to the external
``pump" field which is responsible for the parametric amplification
 of tensor metric fluctuations, normalised to an initial vacuum
fluctuation spectrum.

With respect to the standard inflationary scenario, the
amplification of tensor perturbations in the pre-Big Bang scenario
is strongly enhanced for large comoving wavenumber $k$
\cite{Gasperini:1992pa,Gasperini:1993dp}, in such a way that the
produced background of gravitational waves could in principle be detected in
the near future by various planned experiments
\cite{Gasperini:1993em,Gasperini:1992pa,Brustein:1995ah}.
Arguments in favour of this conclusion arise from the particular
kinematics of string cosmology. First, since the Hubble parameter is
increasing in the pre-Big Bang phase, higher frequency modes will cross
the horizon at higher values of $\vert H \vert$, which implies that the
amplitude of the spectrum increases with frequency. In this context,
the comoving amplitude of perturbations may grow even outside
the  horizon \cite{Abbott:1986cu,Gasperini:1993hu,Brustein:1995kn},
instead of being frozen as happens in standard inflation. Second, the
peak amplitude of the spectrum
(naturally located in the high frequency regime, near the end point of the
spectrum) is normalised so as to match the string scale
\cite{Brustein:1995ah,Brustein:1997ut}, i.e. nearly eight orders of
magnitude above the high frequency normalisation of the standard
scenario \cite{Polarski:1996jg}. Hence, the pre-Big Bang phase could lead
to a very
efficient production of gravitons in the sensitivity range of present
gravitational antenna. In fact such a relic background could be in principle
be detected by the second (planned) generation of interferometric
detectors \cite{Thorne:1997ut,Allen:1997ad}.

To date, extended studies of tensor perturbations
have been mostly performed in the context of the tree-level
effective action, with and without the possible contribution of a
high curvature stringy phase. In all cases the background cosmologies
that have been investigated have been singular, i.e. the  curvature
and dilaton have become singular at some fixed proper time.
There have been a number of suggested ways to tackle
the singularity problem, considering also anisotropic
\cite{Gasperini:1991qy,Gasperini:1999zq} and inhomogeneous
\cite{Gasperini:1992ym,Giovannini:1998cv} backgrounds, or the
presence of a non-local dilaton potential \cite{Gasperini:1993em}.
However, if we confine our attention to homogeneous and isotropic metrics,
and to a local potential, one of the most promising
approaches to the graceful exit problem \cite{Brustein:1994kw}
suggests that the curvature singularities may be cured
by adding higher-order corrections to the string effective action (see for instance
\cite{Antoniadis:1994jc,Rey:1996ad,Gasperini:1997fu,Brustein:1998cv,
Foffa:1999dv,Cartier:1999vk}). These include both higher-derivative
$\alpha'$ corrections, due to finite-size effects, and quantum
corrections, due to the loop expansion in the string coupling parameter.
In the context of cosmological perturbation theory,
it is known \cite{Gasperini:1997up} that such corrections may induce an
additional amplification when the curvature becomes large in string
units. The aim of this paper, therefore, is to discuss the evolution and
the spectrum of tensor perturbations in the context of a class of non-singular
cosmological backgrounds, by taking into account (in the
perturbation equation) the full contribution of those corrections that
are responsible for the regularisation of the background evolution. In
this way it will be possible (for the first time, to the best of our
knowledge) to follow the complete evolution of perturbations, from
the initial vacuum down to the present time, through a continuous
numerical integration of the linearised perturbation equations.

The paper is organised as follows. In Section 2 we recall the
general form of the first order $\alpha'$ corrections arising in
the context of the massless bosonic sector of the
low-energy heterotic string action. We extend
them by including possible one- and two-loop quantum corrections,
and we write down the corresponding background equations. In
Section 3 we compute the linearised tensor perturbation equation,
including such higher-order corrections. In Section 4 we
present a general discussion of the evolution of perturbations in
the presence of higher-order curvature corrections. We then check
the main points of our discussion through a
numerical integration of the perturbation equations for regular
backgrounds, by using an appropriate dilaton potential, or particle
production effects, to stabilise the dilaton in the post-Big Bang era.
We find that the general effect of the higher-order corrections is
to flatten the slope of the high-frequency part of the spectrum, but
the effect is probably too small -- in the class of models discussed in
this paper -- to make the graviton background ``visible" to the planned
advanced detectors.
The main results of this paper are finally summarised in Section 5.

\section{Effective action and background equations}

In the context of the pre-Big Bang scenario, our present FRW
Universe is expected to emerge from the string perturbative vacuum
as a consequence of some process -- for instance, plane wave collisions
\cite{Feinstein:2000ja} -- triggering the gravitational collapse of a
sufficiently large portion of spacetime \cite{Buonanno:1998bi}. The
initial evolution is  driven by the kinetic energy of the dilaton,
and can be described, in the string frame, as inflationary expansion
with growing coupling and curvature. After reaching a maximal scale
controlled by the fundamental string length $\lambda_S$,
$|H| \sim {\cal O}(\lambda_{S}^{-1})$, it is then expected that the
Universe is smoothly connected to the FRW regime, with a constant
dilaton field.

In the early stage of the dilaton-driven inflationary (DDI) period,
strings are propagating in a background of small curvature, and the
fields are weakly coupled, so that the cosmological evolution is
consistently described by the lowest order string effective action
which, by  assuming static internal dimensions, can be written as:
\begin{equation}  \label{e:Low}
\Gamma = \frac{1}{\alpha'}\int d^{4}x \sqrt{-g}\left\{ e^{- \phi}
\left[R + (\partial_{\mu} \phi)^2\right] + {\cal L}_{c} \right\}.
\end{equation}
Here $\alpha'= \lambda_S^2$, we have adopted the convention
$(-,+,+,+)$,
$R^{\mu}_{\;\nu\lambda\rho}=\partial_{\rho}
\Gamma^{\mu}_{\nu\lambda}+\dots$, $R_{\nu\rho} =
R^{\lambda}_{\;\nu\lambda\rho}$ and set our units such that $\hbar = c
= 16 \pi G = 1$. Following \cite{Brustein:1998cv}, we have
introduced an additional lagrangian ${\cal L}_{c}$ allowing for the
inclusion of higher-order corrections to the tree-level effective action,
 a non-perturbative potential -- yet to be determined --
and/or the backreaction of the produced radiation.
Considering this additional lagrangian as an independent source
we shall assume that the associated energy-momentum tensor
is diagonal, homogeneous and isotropic, and can be written in the
perfect fluid form as $T^{\mu}_{\nu} = {\rm diag}(-\rho_{c},p_{c},p_{c},p_{c})$.

The resulting equations of motion for the background are
obtained from the variation of Eq.~(\ref{e:Low}) with respect to the
metric and the dilaton field:
\begin{eqnarray}
6 H^2-6 H \dot{\phi}+ \dot{\phi}^2 &=& e^{\phi} \rho_{c},
\label{e:eq_rho} \\
4\dot{\phi}H -4 \dot{H} - 6 H^2 - \dot{\phi}^2 + 2\ddot{\phi} &=& e^{\phi} p_{c},
\label{e:eq_p}\\
6 \dot{H} + 12 H^2 + \dot{\phi}^2 - 2\ddot{\phi} - 6 \dot{\phi}H &=& e^{\phi}\Delta_{\phi},
\label{e:eq_phi} \\
\dot{\rho_c} + 3 H ( \rho_c + p_c) &=& -\Delta_{\phi}\dot{\phi}.
\label{e:E_constraint}
\end{eqnarray}
A dot represents the differentiation with respect to proper
time $t$ and $H=\dot{a}/a$ is the Hubble parameter.
$\Delta_{\phi}$ arises from the variation of the
additional lagrangian with respect to $\phi$.
Finally, Eq.~(\ref{e:E_constraint}) is the conservation equation,
a direct consequence of Eq.~(\ref{e:eq_rho})--Eq.~(\ref{e:eq_phi}).

To lowest order, the homogeneous and isotropic version of the
pre-Big Bang scenario, without non-local dilaton potential,
inevitably faces a curvature and dilaton singularity at the end of the
DDI phase. However, a singular behaviour is often a late manifestation
of the breakdown of the description, extended beyond its domain of
validity. Hence, the low-energy dynamics should be supplemented by
corrections, in order to give a reliable description of  the high
curvature regime, expected to be crossed before reaching the
FRW universe.  Such corrections are generally of two types:
tree-level $\alpha'$ corrections, referring to the finite size of the
string, and quantum corrections, resulting from a more conventional
loop expansion in the string coupling parameter.

It has been shown in \cite{Gasperini:1997fu} that the higher
curvature $\alpha'$ terms, arising from the
expansion around the point-particle limit, may eventually
stabilise the evolution of the Universe into a de-Sitter like regime of
constant curvature $H \sim {\cal O}(\lambda^{-1}_{S})$ and linearly
growing (in cosmic time) dilaton. In this paper we will
consider the most general tree-level $\alpha'$ corrections to the
gravi-dilaton string action up to fourth-order in derivatives
(see \cite{Cartier:1999vk} for a detailed analysis),
which can be written in the form \cite{Metsaev:1987zx}:
\begin{equation}
{\cal L}_{\alpha'} = \alpha' \lambda_{0} e^{-\phi} \left\{ c_1 R^{2}_{GB} + c_2
G^{\mu\nu}\partial_{\mu} \phi
\partial_{\nu} \phi + c_3 \Box \phi (\partial_{\mu} \phi)^{2} + c_4
(\partial_{\mu} \phi)^{4} \right\}, \label{e:Class}
\end{equation}
\begin{equation}
c_1=-1, \hspace{1cm} c_2 + 2 (c_3 + c_4) = - 2 c_1. \label{e:constraint}
\end{equation}
The constraint Eq.~(\ref{e:constraint}) on the coefficients
is required so that the action reproduces the usual string scattering
amplitudes \cite{Metsaev:1987zx}.
The parameter $\lambda_{0}$ allows us to move between
different string models, and we will later use
$\lambda_{0}=-1/4$ to agree with previous studies of the heterotic
string \cite{Gasperini:1997fu}. Finally, $R^{2}_{GB} =
R_{\mu\nu\lambda\rho} R^{\mu\nu\lambda\rho} -4
R_{\mu\nu}R^{\mu\nu} + R^2$ is the Gauss-Bonnet combination,
guaranteeing the absence of derivatives higher than two in the
equations of motion.

Another conventional expansion of the string effective action
is controlled by  the string coupling parameter,
$g^2_{string} = e^{\phi}$. Initially, $g^2_{string} \ll1$. However,
the  growth of the dilaton during
the DDI era implies that the late  evolution of the background
could be dominated by quantum loop corrections, which  will eventually
force the universe to escape from the fixed point
determined by the $\alpha'$ corrections \cite{Gasperini:1997fu},
for a smooth connection to the post-Big Bang regime.
Unfortunately, there is as yet no
definitive calculation of the full loop expansion in string theory, so we
are left to speculate on plausible terms that will eventually make up
the loop contribution. Multiplying each term of the tree-level
$\alpha'$ correction by a suitable power of the string coupling is the
approach which has already met with some
success \cite{Brustein:1998cv,Brustein:1999yq,Cartier:1999vk},
and which we shall adopt also in this paper.
Since the quantum corrections are not formally derived from
an explicit computation, we shall allow for different
coefficients $d_i$ ($e_i$) at one-loop (two-loop) in the string coupling,
replacing the coefficients $c_i$ of the tree-level action,
 where $d_i$  and $e_i$ are not  necessarily subject to the
constraint Eq.~(\ref{e:constraint}). With such assumptions,
the effective lagrangian density ${\cal L}_{c}$ of Eq.~(\ref{e:Low}),
which is in general the sum of the tree-level $\alpha'$ and loop corrections,
 ${\cal L}_{c} = {\cal L}_{\alpha'} + {\cal L}_{g}$,
 in the present case will take the form
\begin{equation}
{\cal L}_{c} =  {\cal L}_{\alpha'}(c_i) + A e^{\phi} {\cal
L}_{\alpha'}(d_i) + B e^{2\phi} {\cal L}_{\alpha'}(e_i) , \label{e:efflag}
\end{equation}
where ${\cal L}_{\alpha'}(c_i)$ is given by Eq.~(\ref{e:Class}),
with the  constant parameters $A$ and $B$ actually
controlling the onset of the loop corrections.

The very similarity between the first order $\alpha'$ and loop
corrections enables a simple form for their contributions to the
background equations of motion. From now on, we
introduce an additional parameter $n$ in the exponential of the
correction ${\cal L}_{c}$, such that $\alpha'$ corrections correspond to
$n=0$, whereas one- or two-loop corrections correspond to $n=1$ and
$n=2$ respectively. For instance, using the notation $\{\dots\}_{1}$ to
represent the one-loop contribution ($n=1$ inside the brackets),  we
easily get the source terms for the system
Eq.~(\ref{e:eq_rho})--Eq.~(\ref{e:eq_phi}):
\begin{eqnarray}
\rho_{c} &=& \{\rho_{c}\}_{0} + A  \{\rho_{c}\}_{1} + B \{\rho_{c}\}_{2}, \label{e:eq_rho_bis} \\
p_{c} &=&  \{p_{c}\}_{0} + A \{p_{c}\}_{1} + B \{p_{c}\}_{2}, \label{e:eq_p_bis}\\
\Delta_{\phi} &=&  \{\Delta_{\phi}\}_{0} + A \{\Delta_{\phi}\}_{1}
+ B  \{\Delta_{\phi}\}_{2}, \label{e:eq_phi_bis}
\end{eqnarray}
and the terms in brackets are given respectively by:
\begin{eqnarray*}
\{{\rho}_{c}\}_{n} &=& \alpha' \lambda_{0} \dot{\phi}e^{(n-1)\phi}
\Bigl\{-24 c_1(n-1) H^3 + 9 c_2 \dot{\phi} H^2 + 6 c_3
\dot{\phi}^2 H - \left[ c_3 (n-1) -  3 c_4 \right] \dot{\phi}^3 \Bigr\}, \\
&& \nonumber \\
\{p_{c}\}_{n} &=& \alpha'
\lambda_{0}e^{(n-1)\phi} \Bigl\{\left[8 c_1 (n-1)^2  - 3 c_2 \right]
\dot{\phi}^2 H^2 - 2 c_2(n-1)  \dot{\phi}^3 H
- 2 c_{3} \dot{\phi}^2 \ddot{\phi}\\
&&\hspace{2,1cm}
 + 8 c_1 (n-1)  \ddot{\phi} H^2 + 16 c_1(n-1)
\dot{\phi} H \left[\dot{H} + H^2 \right] \nonumber \\
&&\hspace{2,1cm} - 4 c_2 \dot{\phi} \ddot{\phi} H -2 c_2
\dot{\phi}^2 \dot{H} \nonumber + \left[ c_4 - c_3(n-1)
 \right]\dot{\phi}^4\Bigr\},\\
&& \nonumber \\
\{\Delta_{\phi}\}_{n} &=&  \alpha' \lambda_{0} e^{(n-1)\phi}
\Bigl\{24c_1(n-1) H^2 \left[\dot{H} + H^2 \right] - 3
c_4\dot{\phi}^2 \left[ 4 \ddot{\phi} + 4 \dot{\phi}H + (n-1)
\dot{\phi}^2 \right] \nonumber
\\ && \hspace{2,1cm}+ c_3 \dot{\phi} \left[ (n-1)^2 \dot{\phi}^3 +
4 (n-1) \dot{\phi} \ddot{\phi} - 6 \dot{\phi} \left(\dot{H} + 3
H^2 \right) - 12 \ddot{\phi} H \right]\Bigr\}
\nonumber \\ &&
\hspace{2,1cm}- 3 c_2 \left[2 \ddot{\phi} H^2 + 6 \dot{\phi} H^3 +
4 \dot{\phi} H \dot{H} + (n-1) \dot{\phi}^2 H^2 \right] \Bigr\}. \nonumber
\end{eqnarray*}
Here, of course, for $n=1$ the coefficients $c_i$ are replaced by
$d_i$, while for $n=2$ the  $c_i$ are replaced by $e_i$. Note that
for $n=0$, $\lambda_0=-1/4$, $c_4=-c_1=1$, one easily
recovers the $\alpha'$ corrections used in
\cite{Gasperini:1997fu,Gasperini:1997up}.

In general, the combination of tree-level $\alpha'$ and quantum loop
corrections
does not lead to a fixed value for the dilaton in a finite amount of time.
Although a non-perturbative (supersymmetry breaking) potential \cite{Kaloper:1993mq}
is expected to stabilise the dilaton, here we shall follow \cite{Brustein:1998cv} by
assuming the dilaton is frozen out by radiation, after the transition to
the post-Big Bang regime. To this aim we will introduce ``by hand" the
presence of radiation, coupled to the dilaton and satisfying the
conservation equation
\begin{equation}
\dot{\rho} + 4
H \rho - \frac{1}{2}\Gamma_{\phi}\dot{\phi}^2 = 0 ,
\end{equation}
where $\Gamma_{\phi}$ represents the decay width of the dilaton.

The solutions of the system of
equations Eq.~(\ref{e:eq_rho}) -- Eq.~(\ref{e:eq_phi}), including the
radiation and the quantum corrections, provide the cosmological
gravi-dilaton background in which we will study the propagation of
tensor perturbations. We will consider a spatially flat FRW manifold,
and we will perturb the above equations keeping the dilaton and all
sources fixed, $\delta \phi=0$, $\delta T^{\nu}_{\mu} = 0$.
To lowest order, the tensor fluctuations will obey the usual string
frame perturbation equation \cite{Gasperini:1993dp}, including the
non-minimal coupling to a time-dependent dilaton.
On the other hand, the inclusion of higher-order corrections, needed
to regularise the background, will inevitably lead to a modification of such a
perturbation equation
(see \cite{Starobinsky:1981zc,Mukhanov:1987pv,Amendola:1989vj,Noh:1997da}
for initial studies taking into account higher curvature
contributions to the evolution of perturbations).

\section{Tensor perturbations}
It is well known that gravitational waves, arising from linearised tensor
perturbations, do not couple to pressure and energy density, and hence do not
contribute to classical gravitational instabilities. Nevertheless, they
are of interest  as a specific signature of pre-Big Bang cosmologies since,
as stressed in a number of papers
\cite{Gasperini:1993em,Gasperini:1992pa,Gasperini:1993dp},
the production of high-frequency gravitons is strongly enhanced,
in string cosmology, with respect to the standard inflationary
scenario. In this study we will restrict our attention to tensor
metric perturbations around a $d=3$ spatially flat background,
parameterised by the transverse, trace-free variable $h_{\mu\nu}$,
\begin{equation}
g_{\mu\nu} \rightarrow g_{\mu\nu} + h_{\mu\nu}
\hspace{1cm}\textrm{with}\hspace{0,5cm} \nabla_{\nu}h_{\mu}^{\nu}
= 0 = h_{\mu}^{\mu},
\end{equation}
where $\nabla_{\nu}$ denotes covariant differentiation with
respect to the background metric. Note also that the indices of
$h_{\mu\nu}$ are raised or lowered with the unperturbed metric,
$h_{\mu}^{\nu}= g^{\nu\rho}h_{\mu\rho}$.

\subsection{Linearised equations}
The linear evolution equation for the metric fluctuations can
be obtained by perturbing the action Eq.~(\ref{e:Low})
up to terms quadratic in the tensor variable $h_{\mu\nu}$:
\begin{eqnarray}
\delta^{(2)}\Gamma &=& \frac{1}{\alpha'} \int d^{4}x
\;e^{-\phi} \left[ \delta^{(2)}(\sqrt{-g}R) +
\delta^{(2)}(\sqrt{-g}\partial_{\mu}\phi\partial^{\mu}\phi)\right]
\nonumber \\ &&
+\lambda_{0} \int d^{4}x \left[ \{\delta^{(2)}{\cal L}\}_{0} +
A\{\delta^{(2)}{\cal L}\}_{1} +B\{\delta^{(2)}{\cal L}\}_{2}
\right],
\end{eqnarray}
where the integrand of the $n$-correction is given by:
\begin{eqnarray}
\{\delta^{(2)}{\cal L}\}_{n} &=& e^{(n-1)\phi} \biggl\{
\delta^{(2)}(c_1\sqrt{-g}R^{2}_{GB}) + \delta^{(2)}\left(c_2\sqrt{-g}\left[
R^{\mu\nu}-\frac{1}{2}g^{\mu\nu}R\right] \partial_{\mu} \phi
\partial_{\nu} \phi\right)
\nonumber \\ && \hspace{2cm}+
\delta^{(2)}\left(c_3\sqrt{-g}\Box \phi (\partial_{\mu}
\phi)^{2}\right) + \delta^{(2)}\left(c_4\sqrt{-g}(\partial_{\mu}
\phi)^{4} \right)\biggr\}.
\end{eqnarray}

Following \cite{Gasperini:1997up}, it is convenient to use
the synchronous gauge where
\begin{eqnarray}
g_{00} = -1, & ~~g_{0i} = 0, ~~& ~~g_{ij} = a^2\delta_{ij}, \nonumber \\
h_{00} = 0,\hspace{0,3cm} & ~~h_{0i} = 0, ~~& g^{ij}h_{ij}=0
\hspace{0,5cm}\textrm{and}\hspace{0,5cm}\partial_{j}h^{j}_{\;i} =
0.
\end{eqnarray}
This enables us to write the perturbed action
as a quadratic form depending on the first and second derivatives
of the symmetric, trace-free matrix $h = h^{i}_{\;j}$, with
time-dependent coefficients fixed by the background fields $a(t)$,
$\phi(t)$. Introducing the matrix notation $\textrm{Tr} \; h^2 \equiv
h_{i}^{\;j} h_{j}^{\;i}$,  and using $\nabla^2 = \delta^{ij}\partial_{i}
\partial_{j}$ for the Laplace operator in a flat space, we have
\begin{eqnarray}
\delta^{(2)}\Gamma &=& \frac{1}{\alpha'} \int d^{4}x \;a^{3}
e^{-\phi}\;\textrm{Tr}\;\Biggl\{ \left[\frac{1}{4} \dot{\phi}^2 -
\frac{3}{2} \dot{H} - 3 H^2\right] h^2
\nonumber \\ &&
\hspace{3,5cm}- h\ddot{h} - 4 H h \dot{h} -
\frac{3}{4} \dot{h}^2 + \frac{1}{4} h \frac{\nabla^2}{a^2}h \Biggr\}
\nonumber \\ &&
+\lambda_{0} \int d^{4}x \;\textrm{Tr}\;
\left[ \{\otimes\}_{0} + A\{\otimes\}_{1} +B\{\otimes\}_{2} \right],
\end{eqnarray}
where the integrand of the $n$-correction is given by:
\begin{eqnarray*}
\{\otimes\}_{n} &=&a^{3} e^{(n-1)\phi} \Biggl\{\left[-6 c_{1} H^2
\left(\dot{H}+H^2\right) - \frac{1}{4} \dot{\phi}^2 \left(3 c_{2}
H^2 + c_{3} \ddot{\phi} + 3 c_3 \dot{\phi} H + c_{4} \dot{\phi}^2
\right) \right]  h^{2}  \\
&&\hspace{1,8cm} - \left[ c_{1} \dot{H}
+7 c_1 H^2 + \frac{1}{8}c_{2}\dot{\phi}^2\right] \dot{h}^2 - 4
c_{1} H^2 h \ddot{h} - 2 c_{1} H \dot{h} \ddot{h}
\nonumber \\&&\hspace{1,8cm}
- \left[ 8 c_{1} H (\dot{H} + 2 H^2) + c_{2}
\dot{\phi}^2 H + \frac{1}{2} c_{3} \dot{\phi}^3 \right] h \dot{h} +
2 c_{1} \dot{h} \frac{\nabla^2}{a^2}\dot{h}
\nonumber \\&&\hspace{1,8cm}
+ \left[c_{1} H^{2} + c_{1}\dot{H} + \frac{1}{8}
c_{2} \dot{\phi}^2 \right] h\frac{\nabla^2}{a^2}h + 4 c_1 H
\dot{h}\frac{\nabla^2}{a^2}h + 2 c_1
\ddot{h}\frac{\nabla^2}{a^2}h\Biggr\}. \nonumber
\end{eqnarray*}
Here, and in what follows, the coefficients $c_i$ are to be replaced
by $d_i$ and $e_i$ for $n=1$ and $n=2$, respectively. Again,
 for $n=0$, $\lambda_0=-1/4$, $c_4=-c_1=1$, one  recovers
the first order $\alpha'$ corrections discussed in
\cite{Gasperini:1997up}.

We can now integrate by parts all the terms with more than two partial
derivatives acting on $h$, as well as the terms in $h\dot{h}$ and
$h\ddot{h}$. This will drop all terms with more than two
derivatives thanks to the Gauss-Bonnet combination, leading to
\begin{eqnarray}
\delta^{(2)}\Gamma &=& \frac{1}{\alpha'} \int d^{4}x \;a^3
e^{-\phi}\; \textrm{Tr}\; \Biggl\{ \left(\frac{1}{2} \ddot{\phi} -
\frac{1}{4} \dot{\phi}^2 - \dot{H} - \frac{3}{2} H^2 + \dot{\phi}
H \right) h^2 \nonumber
\\ && \hspace{3,5cm} +\frac{1}{4}
\dot{h}^2 + \frac{1}{4} h\frac{\nabla^2}{a^2}h \Biggr\} \nonumber
\\ &&+\lambda_{0} \int d^{4}x \;\textrm{Tr}\; \left[ \{\otimes\}_{0} +
A\{\otimes\}_{1} +B\{\otimes\}_{2} \right], \label{e:eq_2_var}
\end{eqnarray}
where we have
\begin{eqnarray*}
\{\otimes\}_{n} &=&a^{3} e^{(n-1)\phi} \Biggl\{-\frac{1}{4}h^2 \Bigl( \left[
8 c_1(n-1)^2- 3 c_2\right] \dot{\phi}^2 H^2 - 2 c_2 (n-1) \dot{\phi}^3 H \\
&& \hspace{1,9cm}
+ \left[c_4 - c_3 (n-1)\right] \dot{\phi}^4
-2 c_3 \dot{\phi}^2 \ddot{\phi} +8 c_1 (n-1) \ddot{\phi} H^2
\nonumber\\ &&\hspace{1,9cm}
+16 c_1 (n-1) \dot{\phi} H \left[\dot{H} +H^2\right]
 -4 c_2 \dot{\phi} \ddot{\phi} H -2 c_2 \dot{\phi}^2 \dot{H} \Bigr)
 \nonumber \\
 &&\hspace{1,9cm}+ h\frac{\nabla^2}{a^2}h \left( \left[ c_1 (n-1)^2 + \frac{1}{8}
c_2\right] \dot{\phi}^2 + c_1(n-1) \ddot{\phi}\right)
\nonumber \\
&&\hspace{1,9cm}+ \dot{h}^2 \left( c_1 (n-1) \dot{\phi} H -
\frac{1}{8} c_2 \dot{\phi}^2 \right)\Biggr\}.
\end{eqnarray*}
We stress that all tree-level $\alpha'$ and quantum loop corrections
disappear in the limit of a constant dilaton field, since in that case the
contribution of the Gauss-Bonnet term corresponds a topological invariant.
This ensures that the resulting perturbation equation reduces to the
Klein-Gordon form, typical of a  minimally coupled scalar,
as we enter the frozen dilaton, post-Big Bang  regime.

The rather complicated form of Eq.~(\ref{e:eq_2_var}) can be simplified
as the coefficient of the $h^2$ term vanishes identically,
thanks to the background equation Eq.~(\ref{e:eq_p}).
By decomposing the matrix $h^{i}_{\;j}$ into the two physical
polarisation modes of tensor perturbations, $h_{+}$ and $h_{\times}$,
\begin{equation}
\textrm{Tr} \; h^2 \equiv h^{i}_{j}h^{j}_{i} = 2 (h_{+}^2 +
h_{\times}^2),
\end{equation}
we can finally write the action, for each polarisation mode
$h(x,t)$, as
\begin{eqnarray}
\delta^{(2)}\Gamma_{h} &=& \frac{1}{2\alpha'} \int d^{4}x \;a^3
e^{-\phi} \Bigl\{ \dot{h}^2 \left( 1 + \{\oplus\}_{0} + A  \{\oplus\}_{1} +
B\{\oplus\}_{2} \right)
\nonumber \\ &&\hspace{3,1cm} +
h\frac{\nabla^2}{a^2}h \left(1 + \{\otimes\}_{0} + A  \{\otimes\}_{1} +
B\{\otimes\}_{2} \right) \Bigr\}, \label{e:action_pol}
\end{eqnarray}
with
\begin{eqnarray*}
\{\oplus\}_{n} &=& \alpha' \lambda_0 e^{n\phi} \left[ 4 c_1 (n-1)
\dot{\phi} H -\frac{1}{2} c_2 \dot{\phi}^2 \right],
\nonumber \\
\{\otimes\}_{n} &=& \alpha' \lambda_0 e^{n\phi}  \left[4 c_1 (n-1)^2
\dot{\phi}^2 + \frac{1}{2} c_2 \dot{\phi}^2 + 4 c_1 (n-1)
\ddot{\phi} \right],
\end{eqnarray*}
where $h$ is now a scalar variable standing for either one of the
two polarisation amplitudes $h_{+}$, $h_{\times}$. The variation
of the action with respect to $h$ gives then the modified
perturbation equation:
\begin{eqnarray}
0 &=&   \left[ 1 + \{\oplus\}_{0} + A\{\oplus\}_{1} + B \{\oplus
\}_{2}\right]  \ddot{h} \nonumber\\ &&+  \left[ 3H - \dot{\phi} +
\{\otimes\}_{0} + A\{\otimes\}_{1} + B \{\otimes\}_{2}\right] \dot{h}
\nonumber
\\ && - \left[ 1 + \{\ominus\}_{0} + A\{\ominus\}_{1} + B \{\ominus
\}_{2}\right] \frac{\nabla^2}{a^2} h, \label{e:eq_pol}
\end{eqnarray}
with
\begin{eqnarray*}
\{\oplus\}_{n} &=& \alpha' \lambda_0 e^{n\phi} \left\{4 c_1 (n-1)
\dot{\phi} H - \frac{1}{2} c_2 \dot{\phi}^2 \right\},
\nonumber \\
\{\otimes\}_{n} &=& \alpha' \lambda_0 e^{n\phi} \Biggl\{ 4 c_1 (n-1)
\left(\ddot{\phi} H + \dot{\phi} \dot{H} \right) - c_2 \dot{\phi}
\ddot{\phi}-\frac{1}{2} c_2 (n-1) \dot{\phi}^3
\nonumber \\ &&
\hspace{1,6cm} + 12 c_1 (n-1) \dot{\phi} H^2 + \left(4 c_1 (n-1)^2
- \frac{3}{2} c_2 \right) \dot{\phi}^2 H \Biggr\},
\nonumber \\
\{\ominus\}_{n} &=& \alpha' \lambda_0 e^{n\phi} \left\{4 c_1 (n-1)^2
\dot{\phi}^2 + \frac{1}{2} c_2 \dot{\phi}^2 + 4 c_1 (n-1)
\ddot{\phi} \right\}.\nonumber
\end{eqnarray*}
Eq.~(\ref{e:eq_pol}) controls the time evolution of the Fourier
components $h_{k}$ of the two polarisation modes, and is valid
even during the high curvature regime connecting the pre-Big Bang branch to our
present FRW universe,  since it incorporates the contributions of the
higher-order corrections needed to regularise the background evolution.

\subsection{Quantisation in conformal time}
We shall now focus on the generation of gravitational waves in the
context of the pre-Big Bang scenario. To normalise the graviton
spectrum to the quantum fluctuations of the vacuum
we need the canonical variable that diagonalises the
perturbed action, and that represents in this case the normal
modes of tensor oscillations of our gravi-dilaton background. To
this purpose we note that introducing the conformal time
coordinate $\eta$, defined by $a = dt / d\eta$, the action
Eq.~(\ref{e:action_pol}) can be re-written
\begin{equation}
\delta^{(2)}\Gamma_{h} = \frac{1}{2\alpha'} \int d^{3}x d\eta
\left\{ z^{2} {h'}^{2} + y^{2} h \nabla^{2} h\right\},
\end{equation}
where a prime denotes differentiation with respect to the
conformal time, and
\begin{eqnarray}
y^{2}(\eta) &=& e^{-\phi} \left[a^2 + \{\oplus\}_{0} + A\{\oplus\}_{1} +
B \{\oplus\}_{2} \right], \label{e:y_gen} \\
z^{2}(\eta) &=& e^{-\phi}
\left[a^2 + \{\otimes\}_{0} + A\{\otimes\}_{1} + B \{\otimes\}_{2} \right], \label{e:z_gen}
\end{eqnarray}
with
\begin{eqnarray}
\{\oplus\}_{n} &=& \alpha' \lambda_0 e^{n\phi} \left\{ 4 c_1 (n-1)^2
{\phi'}^2 + \frac{1}{2} c_2 {\phi'}^2 + 4 c_1 (n-1) \left( \phi''
- \frac{a' \phi'}{a} \right)\right\}, \nonumber \\
\{\otimes\}_{n} &=&
\alpha' \lambda_0 e^{n\phi} \left\{ 4 c_1 (n-1) \frac{a' \phi'}{a}
- \frac{1}{2}c_2 {\phi'}^2\right\}.\nonumber
\end{eqnarray}
Introducing the canonical variable $\psi = z h$ enables us to
diagonalise the kinetic part of the action:
\begin{equation}
\delta^{(2)}\Gamma_{h} = \frac{1}{2\alpha'} \int d^{3}x d\eta
\left\{ {\psi'}^{2} + \frac{z''}{z} \psi^2 + \frac{y^2}{z^2}\psi
\nabla^2\psi \right\}.
\end{equation}
For each Fourier mode, $\psi_{k} = z h_{k}$, we can
now obtain from the perturbed action a
linearised wave equation in terms of the eigenstates of the
Laplace operator, $\nabla^2 \psi_{k} = - k^{2}  \psi_{k} $ which
takes the explicit form:
\begin{equation}
\psi_{k}'' + \left\{ k^2\frac{y^2}{z^2} - V(\eta)\right\} \psi_{k} = 0,
~~~~~~V(\eta) = \frac{z''}{z}\label{e:psi_full}.
\end{equation}
This linear perturbation equation is the major result of the first
part of this paper, since it encodes
(through Eq.~(\ref{e:y_gen}) -- Eq.~(\ref{e:z_gen})) the full contribution
 of those higher-order corrections which can be used
to regularise the background evolution. We note that the
above equation defines an effective
time-dependent mode, $k y/z$, whereas the effective potential is
determined by the evolution of the generalised pump field
$z(\eta)$. These new features of our generalised perturbation
equation will be discussed in detail in the next section.

The graviton spectrum of pre-Big Bang models generated
in the context of the lowest-order perturbation equation, with and
without the contribution of a high curvature string phase, has
been discussed extensively in the literature (see for example
\cite{Gasperini:1994xg,Brustein:1995kn,Brustein:1995ah,Buonanno:1997xc,
Buonanno:1998zk,Gasperini:1999av}). In the remaining part of this
section we shall briefly recall some general results
of these studies, so as to highlight the modifications induced by the higher-order
corrections in the perturbation equation, and in the spectrum of the relic
pre-Big Bang gravitons.

In the extremely weak coupling and low curvature regime, the
tree-level solutions of the pre-Big Bang scenario are fully adequate to
describe, asymptotically, the initial evolution of the quantum
fluctuations of the gravi-dilaton background. In normalising the
fluctuations in the asymptotic past $(\eta \rightarrow -\infty)$, we
shall thus neglect any correction to the low-energy action,
considering the standard perturbation equation for the Fourier
mode $\psi_{k}(\eta)$:
\begin{equation}
\psi_{k}'' + k^2 \left[ 1- \frac{\tau(\tau-1)}{(-k\eta)^2} \right] \psi_{k} =
0.\label{e:psi_reduced}
\end{equation}
Such an equation describes the evolution of a classical harmonic
oscillator undergoing a parametric amplification driven by the effective
potential $V = \xi''/\xi$, with $\xi \sim (-\eta)^\tau$. By recalling
the (conformal time) expression of the tree-level solutions
\cite{Gasperini:1994xg},
\begin{eqnarray}
a(\eta) = (-\eta)^{\alpha}, &~~~~~&\alpha = (1-\sqrt{3})/2, \label{e:low_a}
\\ \phi(\eta) =  -\gamma \ln (-\eta), &~~~~~&\gamma = \sqrt{3}\label{e:low_phi},
\end{eqnarray}
we obtain the external pump field $\xi =a e^{-\phi/2} =
(-\eta)^\tau,\, \tau=1/2$. Hence, the power-law type of the effective
potential leads to two distinct regimes for the evolution equation
Eq.~(\ref{e:psi_reduced}).

Sub-horizon modes $(|k\eta| \gg 1)$ are freely
oscillating, and no quantum particle production occurs. However, the
non-adiabatic behaviour of modes having left the Hubble radius
$(|k\eta| \ll 1)$ in the dilaton-driven phase, are
described asymptotically by the general solution
\cite{Mukhanov:1992tc}
\begin{equation}\label{h_outside} h_{k} (\eta)
= A_{k} + B_{k} \int^{\eta}_{\eta_{exit}} \frac{d\eta'}{\xi^{2}(\eta')},
\end{equation}
where $\eta_{exit} = k^{-1}$, and $h_{k} = \psi_{k} \;\xi^{-1}$ is the comoving amplitude of
tensor perturbation for the given mode ($A_k$ and $B_k$ indicate the
coefficients of the growing and decaying solutions, respectively).
The amplitude of super-horizon modes is thus frozen, modulo
logarithmic corrections. The overlapping of these two extreme
behaviours is given by the exact solution to the Bessel-type equation
Eq.~(\ref{e:psi_reduced}), which can be expressed in a linear combination
of the first- and second-kind Hankel functions:
\begin{equation} \psi_k
(\eta) = (-k\eta)^{1/2} \left\{ \kappa_{1}  H^{(1)}_{\nu} (-k \eta) +
\kappa_{2} H^{(2)}_{\nu} (-k \eta)\right\}, \label{psi_hankel}
\end{equation}
where $\nu = |1 - 2\tau|/2 = 0$ encompasses the background dynamics.

In the asymptotic past ($\eta \rightarrow -\infty$) any given mode $k$
is well inside the Hubble radius (as $H^{-1}$ blows up as we go
back in time in pre-Big Bang models), and  the two types of
Hankel functions $ H^{(1,2)}_{\nu} (k |\eta|)$  are
oscillating, corresponding  to negative and positive frequency modes
respectively. It is therefore possible to normalise tensor
perturbations to a spectrum of quantum fluctuations of the initial
background. Indeed, the annihilation and creation operators resulting
from the expansion of the $h_{\mu\nu}$ field, with $\psi(k,\eta) = a_{k}
\Psi + a^{\dag}_{k} \Psi^{*}$, obey canonical commutation relations
$[a_{k},a_{k'}] =[a_{k}^{\dag},a_{k'}^{\dag}] = 0$ and
$[a_{k},a_{k'}^{\dag}] = \delta_{k k'}$. For the normalisation to an
initial vacuum state, we choose to restrict to positive energy states,
$\psi_k \rightarrow \frac{1}{\sqrt{2k}} \;e^{-ik\eta}$, hence
$\kappa_{1} = 0$ and $\kappa_{2}=
\sqrt{\frac{\pi}{2}}\,e^{-i\frac{\pi}{4}(1+2\nu)}$.

The  effective potential of the perturbation equation is expected
to grow monotonically during the initial
dilaton-driven regime, up to a maximal value reached during the
string phase, while it is expected to decrease rapidly to zero at the
onset of the standard, radiation dominated phase. Emerging from
the asymptotic past, a typical mode $k$ will thus first oscillate inside
the horizon and then progressively feel the potential barrier. The
process of quantum particle production will take place and carry on
until the background enters the standard FRW-regime. After that,
the effective potential identically vanishes, and Eq.~(\ref{e:psi_reduced})
yields the free oscillating solution:
\begin{equation} \label{e:psi_radiation}
\psi_{k} =\frac{1}{\sqrt{2k}} \left\{c_{+}(k)
e^{-ik\eta}+c_{-}(k) e^{ik\eta}\right\} \hspace{1cm}\eta \rightarrow +
\infty,
\end{equation}
with incoming and outgoing waves. The mean number of produced gravitons is
obtained from the square modulus of the Bogoliubov coefficient,  $|c_{-}(k)|^2$,
using the matching of the tensor perturbation variable $h$ and
$h'$ at the onset of the FRW phase.

The low-energy perturbation equation may
allow for estimating a lower bound on the spectrum of the
produced gravitons \cite{Brustein:1998kq}.  For the tree-level solution with
$\tau=1/2$,  by defining the spectral energy density per logarithmic
interval of frequency,  $\rho_{gw}(k) $, one obtains in particular
\cite{Gasperini:1994xg}:
\begin{equation} \label{e:rho_spectrum}
\rho_{gw}(k) \equiv
\frac{d\rho_{gw}}{d \ln k}= \frac{k^4}{\pi^2}|c_{-}(k)|^2 \simeq
\frac{k_{1}^4}{\pi^2}\left(\frac{k}{k_1}\right)^3,
\end{equation}
where $k_1 \equiv |\eta_1|^{-1}$ refers to the maximal height  of the
effective potential barrier, $|V(\eta_{1})|^{1/2} \simeq |\eta_1|^{-1}$.
We recall that the cubic slope of the spectrum is a direct consequence of the
evolution of the background during the DDI branch.
In \cite{Brustein:1995ah,Buonanno:1997xc,Buonanno:1998zk}, the
authors argued that the contribution of a high curvature string phase
would lead in general to a reduction of the slope of the spectrum of
relic gravitational waves. This was explicitly confirmed for a model
based only on the tree-level $\alpha'$ corrections
\cite{Gasperini:1997up}. However, we stress that the investigation of
the evolution of high frequency modes, hitting the potential barrier
during the high curvature regime, needs to be performed using the full
perturbation equation Eq.~(\ref{e:psi_full}), including also the loop
corrections.

\section{Results}
In this section we highlight the modifications to the
tensor perturbation equations induced by the corrections adopted to
regularise the background evolution. To this purpose, we shall
first restrict the discussion to the tree-level $\alpha'$ corrections;
the reason for this choice is twofold.

First, $\alpha'$ corrections
are derived unambiguously since we require them to reproduce the string
scattering amplitude, as opposed to our lack of precise knowledge for
quantum loop corrections, which forces us to speculate on plausible
terms. Second, if any clear imprint of a ``string phase"
-- like the one introduced in
\cite{Brustein:1995ah,Buonanno:1997xc,Buonanno:1998zk} --
is to be found in
the spectrum of relic gravitons, it should be a direct consequence of
a long enough string phase\footnote{By long string phase, we mean $N \gg
1$, where $N=\ln(a)$ is the number of e-folds.}
characterised by a constant Hubble parameter.
As suggested in \cite{Gasperini:1997fu}, this can be achieved
by assuming that the evolution of the universe is driven
by $\alpha'$ corrections into a fixed point in the $(\dot{\phi},H)$ space.

In the final part of this section
we will comment on the impact of the quantum loop corrections,
and confront our predictions with the results obtained through
exact numerical integrations of the perturbation equation.

\subsection{Tree-level $\alpha'$ corrections}
The lowest-order effective action of string theory provides an
adequate description of the dilaton-driven branch of the pre-Big Bang
scenario. In this weakly coupled and low curvature regime,
no correction is  required either to the effective
potential nor to the wavelength of a given perturbation, and the
evolution equation of each Fourier mode reduces to its usual form
Eq.~(\ref{e:psi_reduced}). However, it is well known that the kinetic energy
of the dilaton necessarily drives an isotropic and homogeneous
background towards a curvature singularity, imposing the need
to supplement the tree-level
action with corrections in order to establish a reliable
 description of the high curvature regime.
The very presence of the tree-level $\alpha'$ corrections
Eq.~(\ref{e:Class}), subject to the constraint Eq.~(\ref{e:constraint}),
may lead in general to a regime of constant Hubble parameter
and a linearly growing dilaton, $H=$ const, $\dot{\phi} > 0$,
which seems to represent an exact conformal solution of string
theory also to all orders in $\alpha'$ \cite{Gasperini:1997fu}.

\begin{figure}[ht]
  \begin{center}
  \includegraphics[width=6cm,height=5cm]{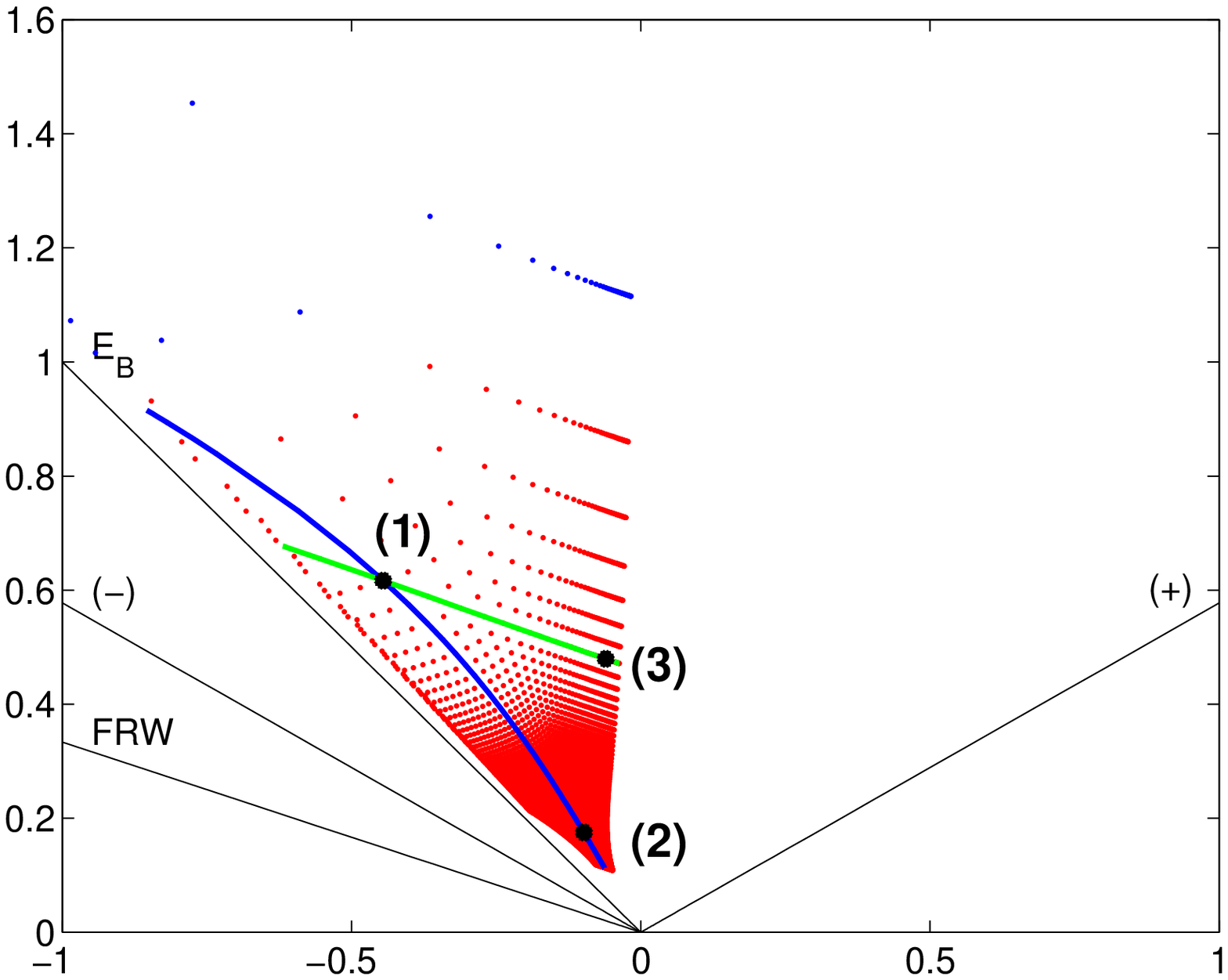}
  \includegraphics[width=6cm,height=5cm]{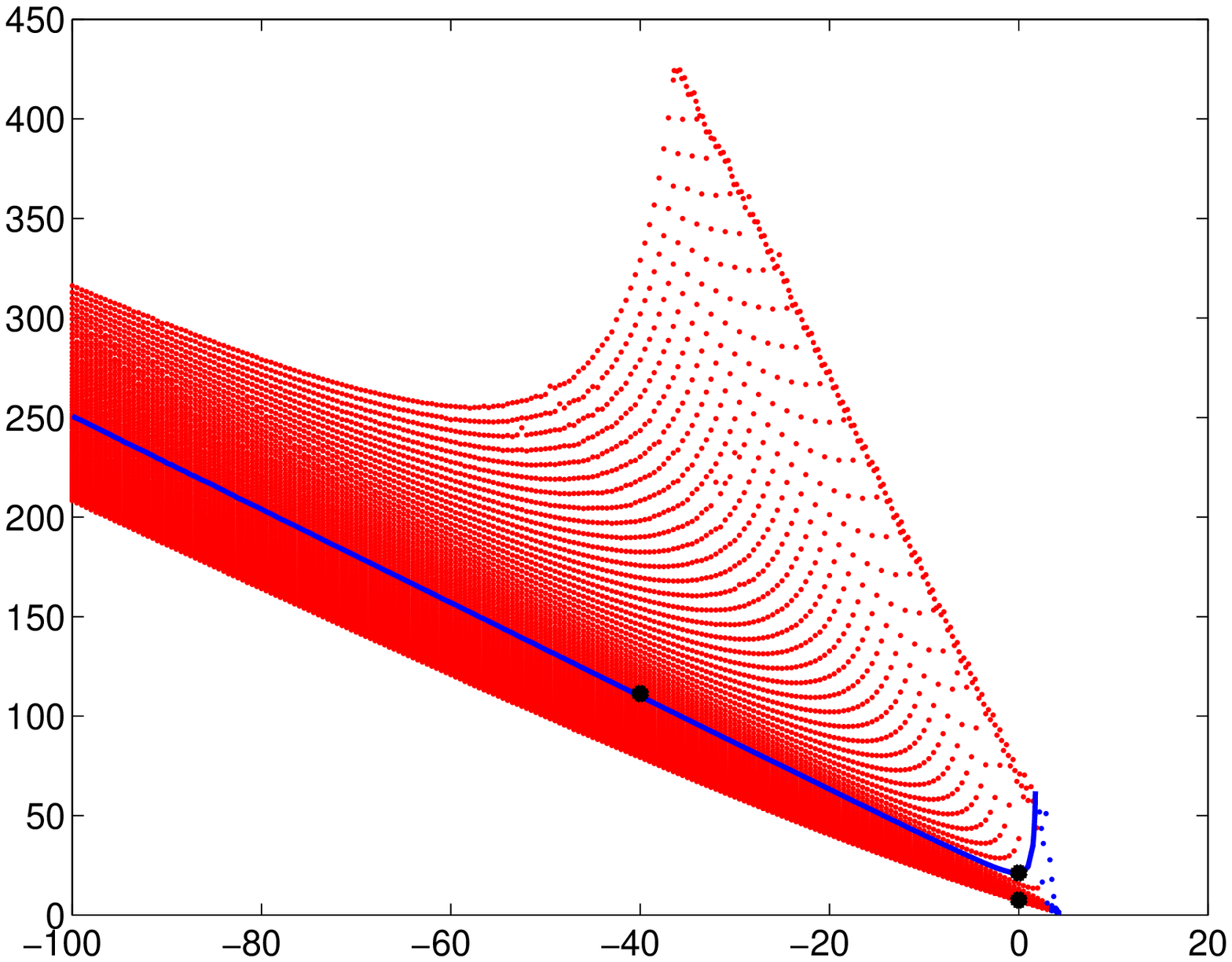}
  \end{center}
  \caption[Location of the attractors and asymptotical frequency shift]
  {\label{f:shift}The figure on the left, from Ref. \cite{Cartier:1999vk},
shows the location in the $(\dot{\bar{\phi}},H)$
space of the fixed points for different choices of the parameters $c_2$
and $c_4,$ and the corresponding ratio $\xi \equiv \dot{\bar{\phi}} /H $.
The blue line shows the $c_2$-dependence of the fixed points for
a constant $c_4=1$, while the green line shows the
$c_4$-dependence for a constant $c_2=0$. The point labeled by (1)
corresponds to $c_2 = 0, \, c_4 = 1$,  with $\xi_{(1)} \simeq -0.72$.
The point labeled by (2)
corresponds to $c_2 = -40, \, c_4 = 1$,  with $\xi_{(2)} \simeq -0.56$.
The point labeled by (3)
corresponds to $c_2 = 0,   \, c_4 = 15$, with $\xi_{(3)} \simeq -0.13$.
On the right, we show the tree-level $\alpha'$ asymptotic shift $c$ given
by Eq.~(\ref{e:classical_shift}) as a function of the parameter $c_2$
and of the fixed point distribution.
The blue line shows the evolution of the shift for a constant $c_4=1$.}
\end{figure}

However, at any given finite order, we recall that not all
sets of coefficients of the truncated action lead to such a good
fixed point in the $(\dot{\phi},H)$ plane (see \cite{Cartier:1999vk}
for a detailed analysis). We shall thus restrict the analysis to the region
of parameters which in principle may allow for  a graceful exit,
when loop corrections are included. In the $(c_2,c_4)$ parameter space,
such a region is delimited on the one hand by $c_2 \leq 38/9$, referring
to a change of sign of the time derivative of the
shifted dilaton, $\dot{\overline\phi}= \dot{\phi} -3H$
\cite{Brustein:1997ny}. On the other hand, tree-level $\alpha'$
corrections, truncated to first order,  are not expected to violate
the null energy condition, and thus cannot lead in principle to fixed
points located after the Einstein bounce \cite{Brustein:1997ny} (i.e,
the bounce of the scale factor in the Einstein frame). This implies
$c_4 \geq c_2/2$ for  $-\infty < c_2 < 4$, to a first approximation.
The location of such fixed points, satisfying the above constraints,
can be seen in Fig.~\ref{f:shift}.

\subsection{Effective frequency shift}
To describe the evolution of the high frequency modes,
which are expected to leave the Hubble radius during the high curvature regime,
we must use the perturbation equation Eq.~(\ref{e:psi_full}),
including those corrections required to regularise the background
evolution. As mentioned in Section $3.2$, the incorporation of the
higher-order corrections in the perturbation equation implies a
time-dependence of  the effective wavenumber $k$. Indeed,
Eq.~(\ref{e:psi_full}) can be conveniently rewritten
\begin{equation}
\psi_{k}'' + \left\{ k^2 [1+c(\eta)] - V(\eta)\right\} \psi_{k} = 0,
\label{41}
\end{equation}
where we have introduced the effective shift in the comoving
frequency $c \equiv (y^2-z^2)/z^2$ which,  at next-to-leading order in
the string tension expansion, reads
\begin{equation}
\label{e:classical_shift} c = \frac{2\alpha'\left\{
(4-c_2){\phi'}^2 - 4 (\phi''- 2\frac{a'\phi'}{a})\right\}} {8
a^2-\alpha'\phi'\left\{8\frac{a'}{a}-c_2 \phi'\right\}}.
\label{1}
\end{equation}

In the weak coupling regime, and for small curvature
$(\alpha' \rightarrow 0)$, the shift in the frequency
is zero, hence no modification to the DDI cubic branch
of the relic gravitational spectrum is expected. However, this
frequency shift will be no longer negligible when
$H \sim {\cal O}(\alpha')$. In general, we find that $c(\eta)$
tends to grow, although a short decrease
(which still satisfies $c(\eta) > -1 $) may happen at the onset
of the string phase. We also observe that the modification
arising in the asymptotic regime, where the curvature is saturated
by a linearly growing dilaton, corresponds to a constant shift of
the comoving frequency, as first suggested in \cite{Gasperini:1997up}.
In that case, by setting $\dot{\phi} = {\phi'}/a \equiv \tilde{\phi}$
and $H= a'/a^2  \equiv \tilde{H}$, we can rewrite the asymptotic version
of Eq.~(\ref{1}) as
\begin{equation}
c = \frac{2\alpha'\tilde{\phi} \left\{ (4 - c_2) \tilde{\phi} + 4
\tilde{H} \right\}}{8 + \alpha'\tilde{\phi}\left\{ c_2
\tilde{\phi} - 8 \tilde{H} \right\}}\label{classical_shift}.
\end{equation}
During such an intermediate string phase the external potential
reduces to its tree-level version, namely $V(\eta) = \xi''/\xi, \quad \xi
=  a e^{-\phi/2}$, and the enhancement in
 frequency associated to Eq.~(\ref{e:classical_shift}) is the same for
all modes, since the shift $c$ does not depend on the wavenumber $k$.

In Fig.~\ref{f:shift} we have represented the effective shift as a
function of the parameter $c_2$, highlighting the $c_2$-dependence by
drawing the curve for $c_4=1$. Although the asymptotic shift cannot
be expressed as a function of an unique variable, it is however possible
to extract its minimal value for the type of $\alpha'$ correction we are
considering. The numerical simulations suggest that the lowest shifts
correspond to fixed points located near the branch change line,
$\dot{\bar{\phi}} \equiv \dot\phi-3H= 0$.
The bound $c_2 = 38/9$ attached to the sign change of
$\dot{\overline\phi}$ implies that, in the limit of large values of
$\tilde{\phi}$, the effective asymptotic shift is always positive, and is
given by $c \geq 10/7$.

Does the frequency shift have any implication on the spectral
distribution? In principle the answer is yes, since higher frequency
modes tend to leave the Hubble radius at larger values of $H$ in the
pre-Big Bang scenario. As a consequence, a given comoving mode will
spend less time under the potential barrier, resulting in a smaller
amplification of the frequency attached to this mode. Although it is
difficult to quantify such an effect, we believe the higher-curvature
corrections amount to an overall rescaling, by a numerical factor of
order unity, of the total energy density of the background, as
already discussed in \cite{Gasperini:1997up}.

Finally, the form of the effective shift Eq.~(\ref{e:classical_shift})
suggests that a background undergoing a non-singular evolution in the
vicinity of $8 -8 \dot{\phi} H + c_2 \dot{\phi}^2=0$ will automatically
imply a rapid growth (if not a divergence) of the amplitude of the
tensor fluctuations. Hence, we can understood the denominator of
Eq.~(\ref{e:classical_shift}) as a source of quantum gravitational
instability, of the type already discussed
in \cite{Kawai:1997mf,Kawai:1999pw}.

\subsection{Full corrections and non-singular evolution}
In recent studies of the graceful exit problem, in the context of
the pre-Big Bang scenario, the
regularisation of the background curvature and the transition to
the post-Big Bang regime, are obtained by including quantum loop
corrections, according to Eq. (\ref{e:efflag}).

In this context, we may note that the tree-level
$\alpha'$ shift Eq.~(\ref{e:classical_shift}) provides an interesting
constraint on the choice of parameters we can use in order to
implement such a graceful exit, since in general the divergence
curve for the high curvature shift is very close to the
condition for the fixed point in the $(\dot{\bar{\phi}},H)$ plane. In
\cite{Cartier:1999vk}, it has been shown that for a constant $c_4$
the fixed point value of the Hubble parameter tends to be
reduced when $c_2$ is decreasing. Simultaneously the location of
the singularity curve also decreases in the $(\dot{\bar{\phi}},H)$ plane,
so that the range of coefficients leading to a non-singular
evolution of the background, and avoiding the instability of
quantum fluctuations, is strongly restricted to values very close to the
simplest case of $\alpha'$ correction, ${\cal L}_{\alpha'} \sim \{ R^2_{GB}
- (\nabla \phi)^4\}$ (as also discussed in \cite{Gasperini:1997fu}).
This is an unexpected, although fortunate constraint on the loop
corrections, which allows restricting the range
of values for the parameters $d_i$ and $e_i$.

In our attempt to characterise the power spectrum of metric perturbations
generated in such a class of non-singular models, we will
thus consider the perturbation equation Eq.~(\ref{41})
with a full shift determined by
\begin{equation}\label{full_shift}
c = \frac{2\alpha'\left\{ {\phi'}^2\left[4-c_2 - A d_2 e^{\phi} - (4e_1+e_2)Be^{2\phi}\right]
    - 4 \left(\phi''- 2\frac{a'\phi'}{a}\right) \left[1+ 4 B e_1 e^{2\phi}\right]\right\}}
{8 a^2-\alpha'\phi'\left\{8\frac{a'}{a}\left[1 + B e_1 e^{2\phi}\right]
-\phi' \left[c_2 + A d_2 e^{\phi} + B e_2 e^{2\phi}\right]\right\}}.
\end{equation}
Here the coefficients of the high curvature, one- and two loop
corrections are respectively $c_i,\;d_i$ and $e_i$, with $c_1=-1$
from the constraint Eq.~(\ref{e:constraint}).

As already stressed, this frequency shift reduces to a constant
in the asymptotic stage of the string phase. The loop corrections
re-establish the time-de\-pen\-den\-ce during the graceful exit,
but its contribution soon becomes negligible at the end of the
phase of accelerated evolution. Indeed,
Eq.~(\ref{e:y_gen}) -- Eq.~(\ref{e:z_gen}) rapidly reduce to their tree-level
expressions $y \simeq z \simeq a e^{-\phi/2}$, hence we expect
$k_{eff} \rightarrow k$ before we enter the radiation-dominated
epoch, where the frequency shift vanishes identically.

\subsection{Observable implications}
We are interested in the generic features of the primordial gravitational waves,
generated by quantum fluctuations of the background metric during
the pre-Big Bang phase, which could be detected by both ground-based detectors
such as LIGO \cite{Allen:1997sw} and VIRGO \cite{Babusci:2000rc}, and
space-based experiments such as LISA \cite{Ungarelli:1999ay} (see also
\cite{Maggiore:1999vm} for a recent review and references therein).
For such a purpose, we use the dimensionless spectral energy
density of perturbations, $\Omega_{gw}(\omega)$, to describe the
background of gravitational radiation. In critical units, the energy
density is defined by
\begin{equation}
\Omega_{gw}(\omega,t) \equiv \frac{1}{\rho_{c}} \frac{d
\rho_{gw}}{d \ln \omega},
\end{equation}
where  $d \rho_{gw}$ is the present energy density in the
stochastic gravitational waves per logarithmic interval of frequency,
 $\omega (t)= k/a(t)$ is the physical (angular) frequency
of the wave, and $\rho_c(t) = 3 M_p^2 H^2(t)/8\pi $ is the critical
energy density required to close the universe.

Following the standard procedure for the computation of the
energy spectrum \cite{Mukhanov:1992tc}, we must determine the
expectation  number of gravitons
per cell of the phase space, $n(\vec{x},\vec{k}) = n_f$, which only
depends on the frequency for an isotropic and stochastic background.
Neglecting $\alpha'$ and loop corrections, both in the background
and perturbation equations, the lowest order result Eq.~(\ref{e:rho_spectrum})
leads to \cite{Gasperini:1994xg}
\begin{equation}\label{e:Omega_tree}
\Omega_{gw}(\omega,t) \simeq \frac{8}{3 \pi}
\frac{\omega_{1}^4}{M_p^2 H^2}
 \left(\frac{\omega}{\omega_1}\right)^3, ~\hspace{1cm}  \omega <
\omega_1
\end{equation}
where $\omega_1 = k_1/a=H_1a_1/a$ is the maximal amplified proper
frequency, associated with the value $H_{1}$ of the Hubble
expansion parameter  at the end of the string phase.

Including higher-order corrections, the relation between the
ultraviolet cutoff of the spectrum and the time of horizon crossing is in
general more complicated, and a precise definition of the maximal
amplified frequency is out of reach without a complete analytic
solution for the background dynamics. However, a reliable estimate
of the slope of the intermediate string branch of the
spectrum  can be obtained by using the
leading features of the string phase.

First, we shall neglect (during the string phase) the
effective  frequency shift in the perturbation equation, whose role
is simply expected to induce an overall rescaling of the amplitude
\cite{Gasperini:1997up}, as already pointed out in Section 4.2.
 Second, we will assume that the
duration of the ``exit phase" is small compared to the intermediate
string phase, and that no dramatic physics take place during the
transition to the standard radiation-dominated epoch
(indeed, in some cases where the exit is catalysed by the loop
corrections, the process is almost instantaneous). This is a big
assumption, of course, which could possibly underestimate
``pre-heating" and ``re-heating" effects. However, the
accuracy of these assumptions can be confirmed with numerical
simulations.

In summary, we suppose that the evolution of the
background consists mainly of three phases:
an initial, low energy dilaton-driven branch, the intermediate
string phase with a constant Hubble parameter ($\tilde H$) and a linearly
growing dilaton ($\dot \phi = \tilde \phi=$ const), and the final
radiation phase with decreasing Hubble parameter
and frozen dilaton. It follows from our assumptions that the
spectral graviton distribution can be correctly estimated through the
low-energy perturbation equation, even for the high frequency branch
of the spectrum.

In that case, it is already well known \cite{Brustein:1995ah} that the
slope of the high-frequency modes, crossing the horizon in the high
curvature, stringy regime, and re-entering in the radiation era, is fully
determined by the fixed point values $\tilde \phi, \tilde H$. Indeed,
during such a string phase, the scale factor undergoes the usual de
Sitter exponential expansion,  while the logarithmic evolution of the
dilaton, in conformal time, is weighted by the ratio $\tilde \phi/\tilde
H$, i.e. $\phi(\eta) \sim - (\tilde \phi /\tilde H) \log (-\eta)+$ const
\cite{Gasperini:1995fm}. By introducing the convenient shifted variable
$\tilde {\overline{\phi}} \equiv \tilde {\phi} - 3 \tilde H$, and referring
the spectrum to a fixed point allowing a subsequent (loop catalysed)
exit, i.e. $\tilde {\overline{\phi}} <0$, one easily finds
\cite{Brustein:1995ah}
\begin{equation}
\Omega_{gw} \sim
\left(\frac{\omega}{\omega_1}\right)^{3-\left|\tilde
{\bar{\phi}}/\tilde H \right|}, ~\hspace{1cm} \omega_s \ll \omega \ll
\omega_1 \label{e:Omega}
\end{equation}
where $\omega_s$ is the limit frequency marking the transition to the
high curvature regime.

In our model of the background evolution, on the other hand, the allowed fixed points
$\tilde \phi, \tilde H$ are located in the ($\dot{\overline \phi}, H$)
plane  between the ``Branch change line'', $\dot{\overline \phi}<0$,
and the ``Einstein Bounce'',
$\dot{\overline \phi}<-H$, see Fig. 1 and \cite{Cartier:1999vk}.
It follows that, in the context of our approximations,
the  spectrum of  tensor fluctuations, at high frequency,
has a slope constrained by
\begin{equation}
\textrm{(Einstein Bounce)} \quad 2 <3 - \left|\tilde {\bar{\phi}}/\tilde H
\right| < 3 \quad \textrm{(Branch change)}.
\label{e:slope}
\end{equation}
In order to check this important analytical result, we analysed
numerical solutions for different choices of coefficients of the
tree-level $\alpha'$ corrections.

\subsection{Numerical results}
The spectral distribution of relic gravitational waves can be obtained
by numerical integration of the perturbation equation
Eq.~(\ref{e:psi_full}), in the background
Eq.~(\ref{e:eq_rho}--\ref{e:E_constraint}). The initial conditions are chosen
in the low curvature and weakly coupled regime and are thus based on
the tree-level solutions, which provide an adequate description of this
phase. We then evolve the system until a given mode re-enters the
Hubble radius, in the late time FRW radiation-dominated epoch, where
we extract the Bogoliubov coefficients by comparing the result of the
simulation and the free-oscillating solution Eq.~(\ref{e:psi_radiation}).

\begin{figure}[ht]
  \begin{center}
  \includegraphics[width=4.5cm,height=3.6cm]{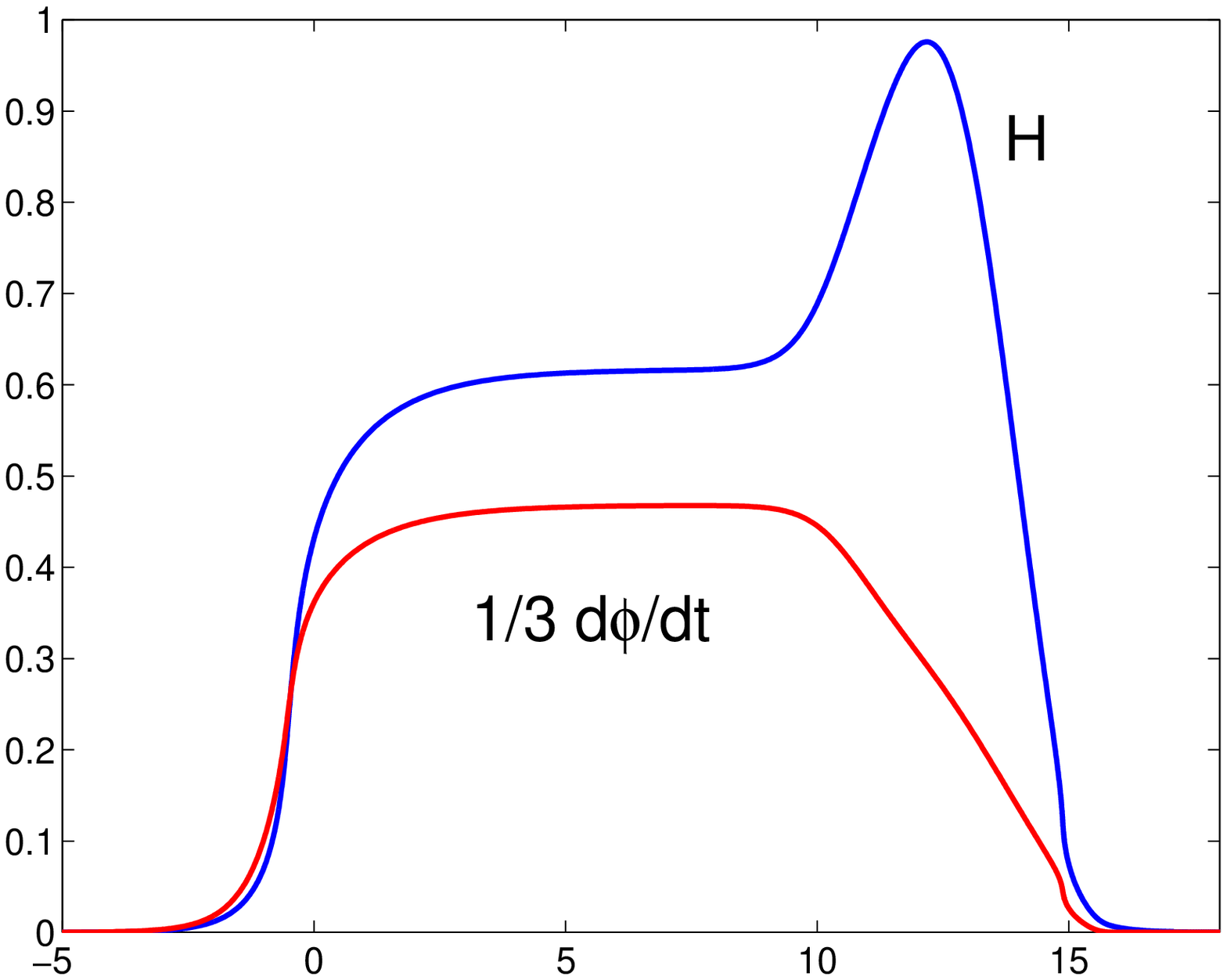}
  \includegraphics[width=4.5cm,height=3.6cm]{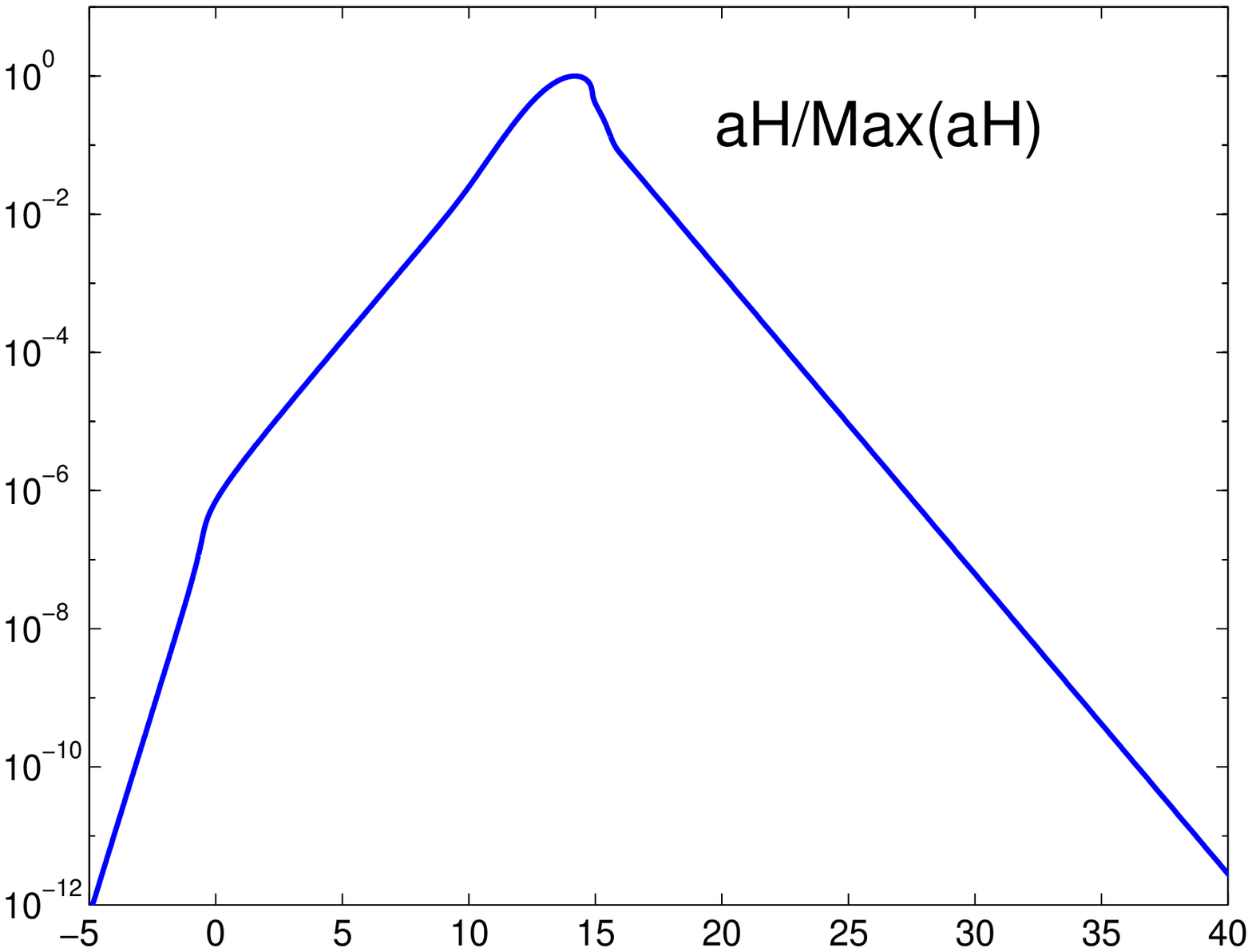}
  \includegraphics[width=4.5cm,height=3.6cm]{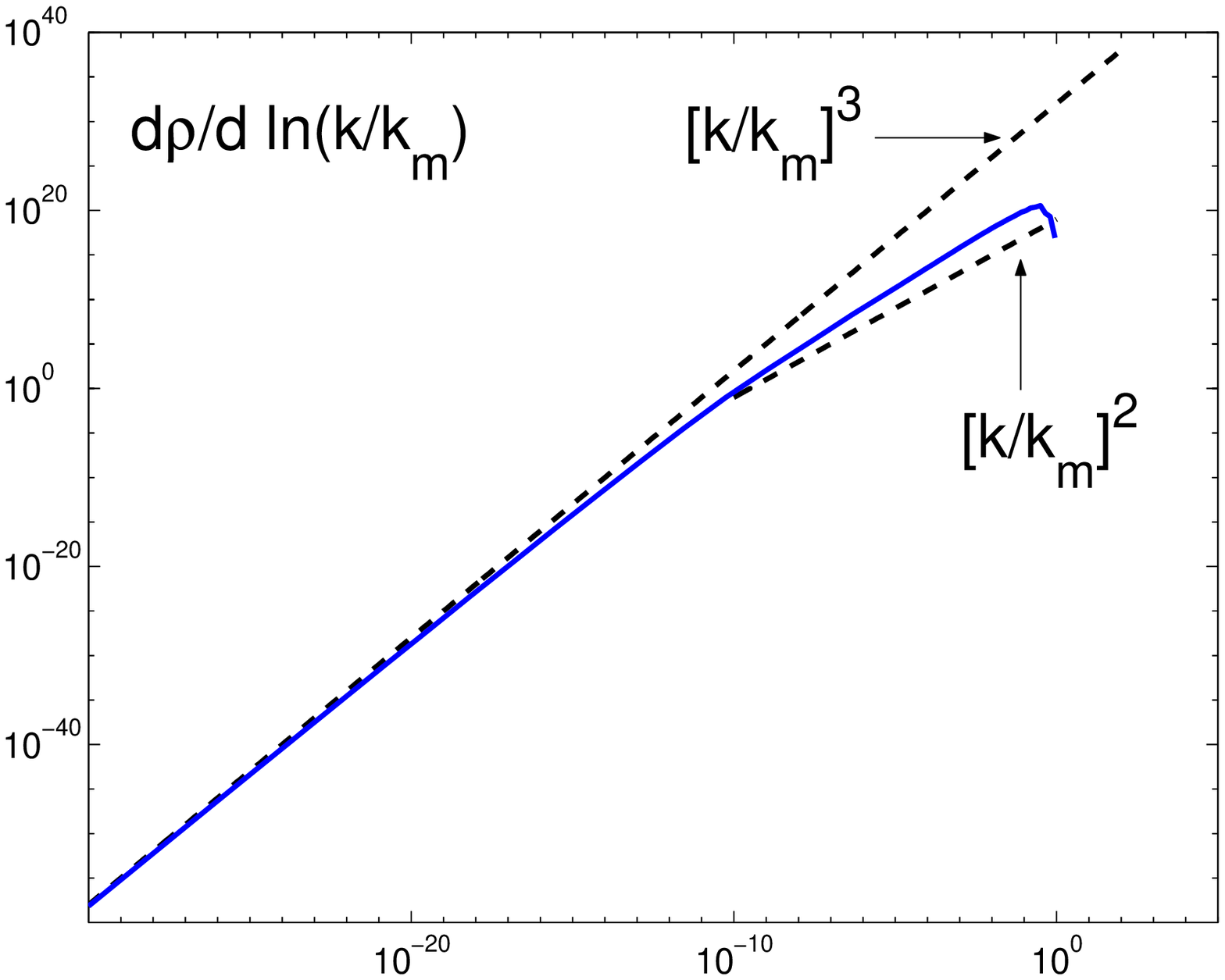}
  \end{center}
  \caption[Spectral distribution of gravitational waves]{\label{f:spectrum}
  The choice of coefficients for the tree-level $\alpha'$ corrections is
  $c_1 =-1,\,c_2 = c_3 = 0$ and $c_4 = 1$ (and the same for $d_i$
  and $e_i$), corresponding to the case $(1)$ in Fig.~\ref{f:shift}.
  The left figure shows a non-singular evolution for the Hubble
  parameter $H=\dot{a}/a$ and for $\dot{\phi}/3$, as a function of the
  number of e-folds, $N=\ln a$.
  The middle figure shows the evolution of $aH/{\rm Max}(aH)$
  as function of $N$. The low-energy, dilaton-driven phase takes
  place approximatively for  $-\infty < N \leq -3$. After a
  short transition, this initial period is followed  by a string phase
  with nearly constant Hubble parameter
  and linearly growing  dilaton, for $2 \leq N \leq 9$.
  After a successful exit triggered by loop corrections,
  the background evolution enters the FRW radiation-dominated phase
  at $N \simeq 16$. Comoving modes leaving the
  Hubble radius during the string phase
  ($10^{-6} \leq k/{\rm Max}(aH) \leq 10^{-2}$)
  are characterised by a slope smaller than $3$ in the spectral
  distribution, as illustrated in
  the right figure which shows $d\rho_k/d\ln(k)$ in units of $k/k_{max}$.
  The upper dashed line corresponds to the cubic slope typical of the
  low frequency part of the spectrum, emerging from the dilaton-driven epoch.}
\end{figure}

To highlight the impact of the time-dependence of the frequency shift,
we first present the spectral distribution in a case where we do not
consider the $\alpha'$ and loop corrections in the perturbation
equation Eq.~(\ref{e:psi_full}).  Fig.~\ref{f:spectrum} illustrates the results
of such a simulation,for the particular model of background
corresponding to $c_1=d_1=e_1=-1$, $c_4=d_4=e_4=1$ (all the other
$c_i,d_i,e_i=0$), $A=1$, $B=-2  \times 10^{-3}$ and $\Gamma_\phi=
5.63 \times 10^{-4}$. We show, in particular,
the  non-singular evolution of $H$ and $\dot\phi$, and
 the evolution of the quantity $aH$, which  enables us
to determine if a given mode leaves the Hubble radius during the DDI
phase  or during the intermediate string phase. Finally, we present the
graviton distribution for such a background, computed from Eq.~(\ref{e:psi_full}),
in units $k/k_{max}$, where $k_{max} = {\rm Max}(aH)$ is a maximal amplified frequency.

The effects of the higher-order corrections are expected
to arise when the curvature scale becomes large in string units. As a
consequence, the low frequency branch of the spectrum is
 unaffected by such corrections,
and remains characterised by a cubic slope. However,
modes leaving the Hubble radius during the intermediate string phase
are strongly affected by the background dynamics. The resulting slope
of the spectrum is found to be reduced, but remains confined
between $2$ and $3$, as expected from Eq.~(\ref{e:slope}).

\begin{figure}[ht]
  \begin{center}
  \includegraphics[width=6cm,height=5cm]{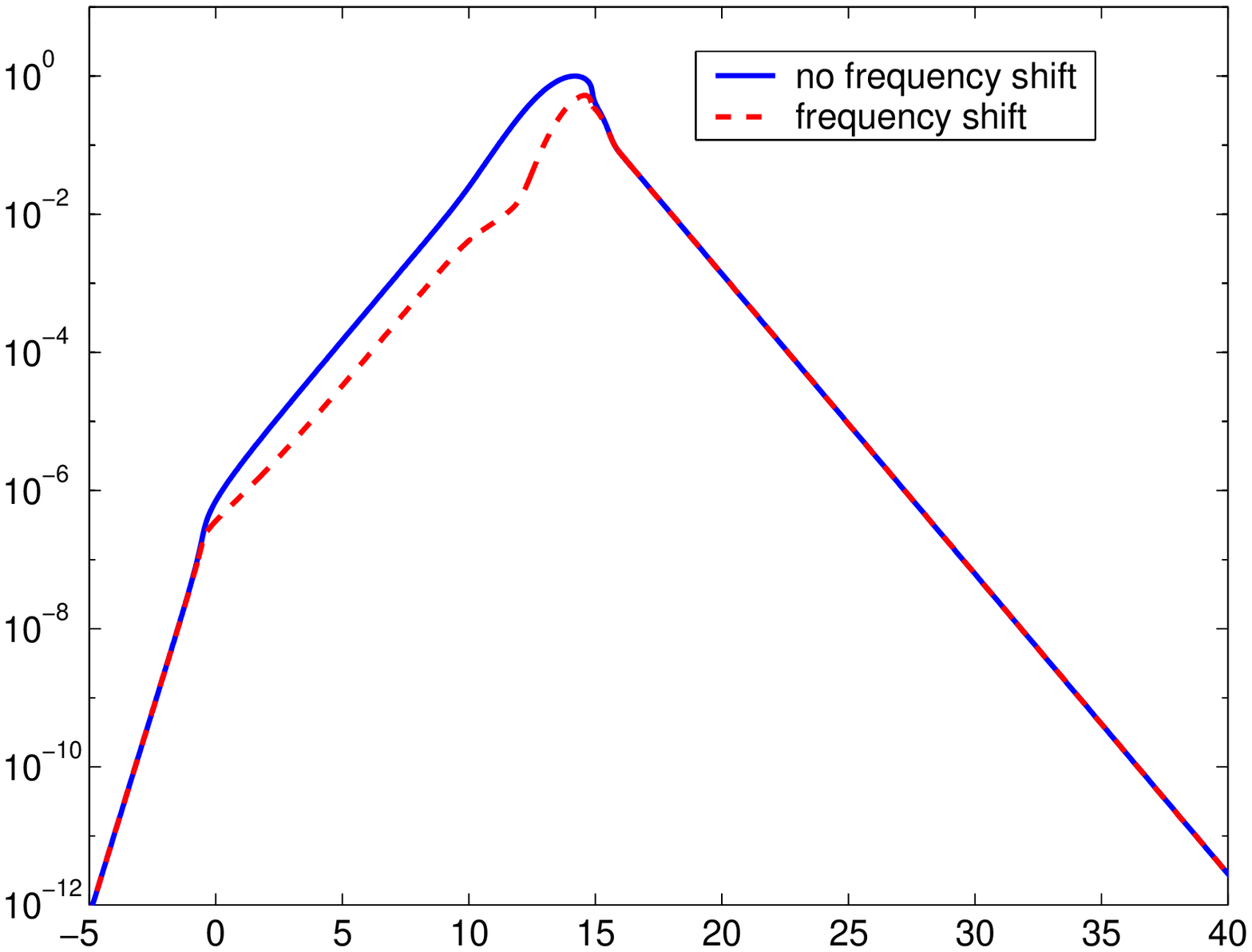}
  \hspace{1cm}
  \includegraphics[width=6cm,height=5cm]{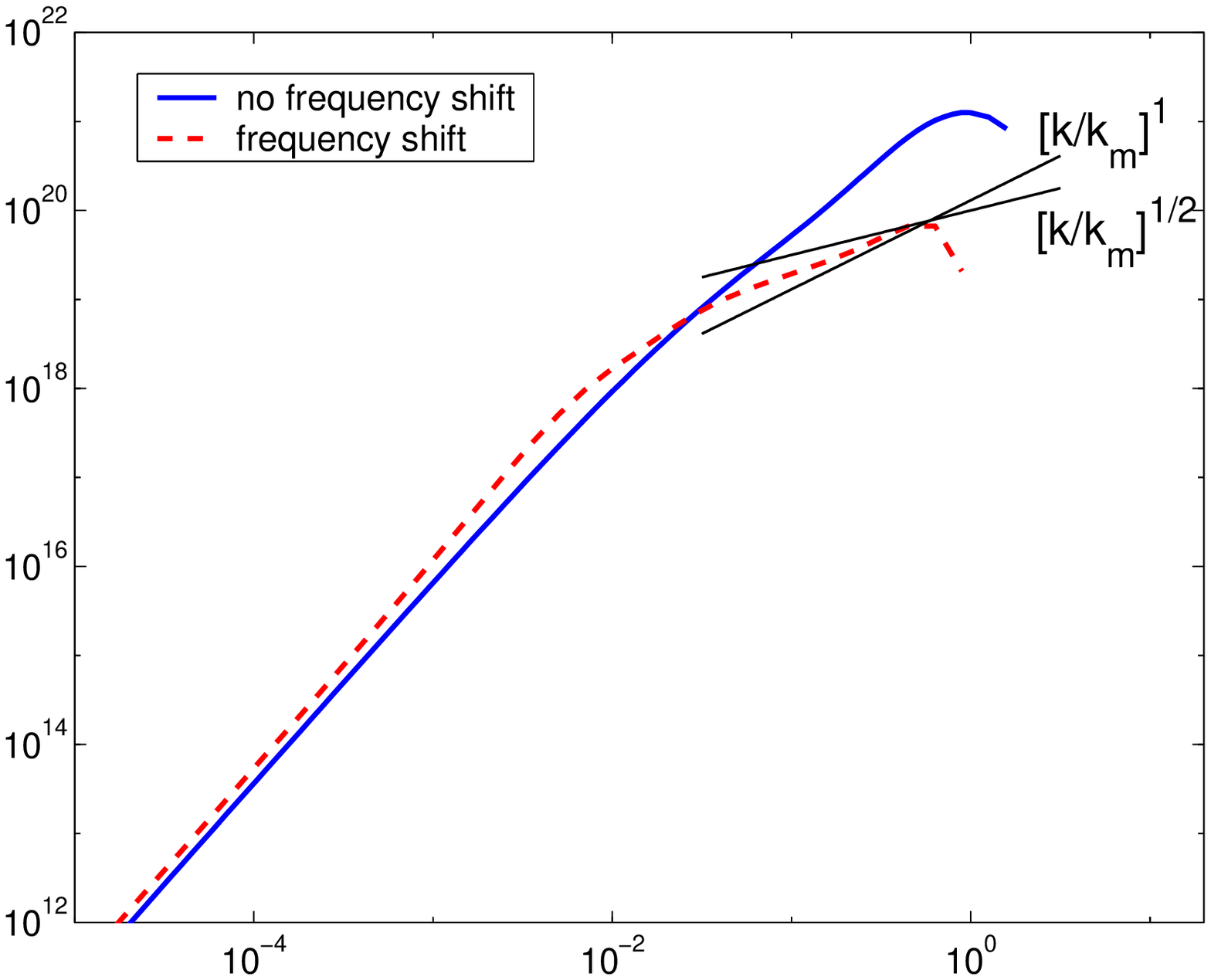}
  \end{center}
  \caption[Spectral distribution of gravitational waves]{\label{f:spectrum2}
  Using the same choice of coefficients as
  in Fig.~\ref{f:spectrum}, we compare the evolution of $aH/{\rm Max}(aH)$ (solid curve)
  and $aH/{\rm Max}\left(aH/\sqrt{1+c}\right)$ (dashed curve)  as a function of $N$.
  When higher-order corrections are included in the  perturbation equation,
  the high frequency cutoff and the duration of the super-horizon epoch are found to be reduced.
  The right hand figure is a magnified image of the end point of the string branch of the
  spectral distribution, which highlights the impact due to a non-trivial frequency shift:
  the significant new feature is the reduction of the high  frequency
  peak and a further reduction of the slope, which may also become
  smaller than two, for comoving modes leaving the Hubble radius
  during the ``exit phase''.}
\end{figure}

Does the inclusion of the time-dependence of the comoving
wavenumber lead to any modification?  We stress that when such a
shift in the frequency is considered, we have $k \sqrt{1+c} = aH$ at
horizon crossing. As a consequence, the shortest scale to leave the
Hubble radius is given by $k_{max} = {\rm
Max}\left(aH/\sqrt{1+c}\right)$, instead of ${\rm Max}(aH)$. This is
illustrated in Fig.~\ref{f:spectrum2}, where we compare the results of the
numerical integration with and without corrections in the perturbation
equation, for identical non-singular background evolutions
(the same model of Fig. 2). No modification appears during the low
energy epoch, where the frequency shift is negligible. However,
higher-order corrections in the perturbation equation, which result in
the time-dependent shift, lead to a reduction of the maximal amplified
frequency and of the time spent on super-horizon scales.

The impact of the time-dependence of the comoving wavenumber $k$
is non-trivial for very high-frequency modes, leaving the Hubble radius
during the graceful exit to the FRW-radiation dominated epoch.
In that case, the result Eq.(\ref{e:Omega}) no longer applies, and the
slope can be smaller than two, as shown in Fig. 3. However, the impact
remains negligible for modes well inside the string branch of the
spectrum i.e. for modes crossing the horizon during the
asymptotic, fixed point regime. In that case, our analytical
approximations for the slope of the spectrum are
confirmed by the numerical simulations. In
Table~\ref{t:slope},  we present the predicted and measured slopes
well inside the string phase, for the
three different choices of coefficients illustrated in
Fig.~\ref{f:shift}. The agreement is striking.

\begin{table}[ht]
\begin{center}
\begin{tabular}{|l|c|c|c|c|}
  \hline
    & \multicolumn{2}{c|}{Coefficients at ${\cal O}(\alpha^\prime)$} & Predicted slope & Measured slope \\
  \cline{2-3}
    & \hspace{0,6cm}$c_2$\hspace{0,6cm} & $c_4$ & $3 - \left|\tilde {\bar{\phi}}/\tilde H \right|$ & (numerical) \\
  \hline
  \hline
  Case $(1)$ &$ 0   $&$ 1  $&$ 2.44 $&$ 2.45 $ \\
  Case $(2)$ &$ -40 $&$ 1  $&$ 2.28 $&$ 2.3  $ \\
  Case $(3)$ &$ 0   $&$ 15 $&$ 2.87 $&$ 2.85 $ \\
  \hline
\end{tabular}
\end{center}
\caption[Predicted and measured slopes of the string branch of the GW spectrum]
{\label{t:slope}Choices of coefficients for the tree-level  $\alpha'$
corrections with $c_1 = -1$ and the remaining coefficient
satisfying $c_3 = -(c_1+c_4+c_2/2)$, according to Eq.~(\ref{e:constraint}).
We have compared the predicted slope determined by the fixed point,
according to Eq.~(\ref{e:Omega}), with the measured slope well inside the string branch of the
spectrum, obtained by numerical integration of Eq.~(\ref{e:psi_full}),
including all corrections.}
\end{table}

It is well known \cite{Gasperini:1999av} that the spectrum of relic
gravitons, in the context of the pre-Big Bang scenario, easily evades
constraints arising from both the CMB anisotropy
at COBE scale \cite{CMB}  and pulsar timing data \cite{Kaspi},
because of its rapid growth at low frequency. Hence,
the peak amplitude of the spectrum is only constrained by the
primordial nucleosynthesis bound \cite{Walker:1991ap},
as well as by primordial black hole production \cite{Copeland:1998gj}.

The  relic graviton background is in general
compatible with these constraints if it is normalised so as to match
the string scale at the high-frequency end point of the spectrum.
In that case one finds \cite{Brustein:1997ut}
that the high frequency peak is
typically of order $\Omega_{gw}(\omega)\simeq 10^{-6}$, for a maximal
amplified frequency $\omega \simeq 10^{11} [{\rm Hz}]$.  Saturating
this high-frequency end point, and using the prediction Eq.~(\ref{e:slope})
for the slope of the  string branch of the spectrum, the energy density
 of relic gravitons from a pre-Big Bang phase could be at
most of ${\cal O}(10^{-15})$ at $\omega \simeq 10^{2} [{\rm Hz}]$,
regardless of the duration of the string phase with constant Hubble
parameter, i.e. far below the sensitivity of the second (planned)  generation of
interferometric gravity wave detectors \cite{Maggiore:1999vm}.
It seems thus difficult to detect the relic gravitons from a
high curvature string phase,  associated to a fixed point of the
truncated effective action -- at least within the model of background
considered in this paper -- unless the exit phase is able to affect a
sufficiently wide band of the high frequency spectrum.

Finally, the numerical integration of Eq.~(\ref{e:psi_full}) enables us to
follow carefully the time evolution of modes, from the DDI epoch up to
the FRW-radiation dominated era. We have
explicitly checked that the amplitude of tensor perturbations is rapidly
frozen in on super-horizon scale, regardless of the
particular higher-order dynamics of the background during the
string phase and the exit epoch.

\section{Conclusion}
In this paper we have discussed the evolution of tensor perturbations in a
class of non-singular cosmological models based on the gravi-dilaton
string effective action, expanded up to first order in $\alpha'$, and
including one-loop and two-loop corrections in the string coupling
parameter.

The coefficients of the loop expansion have been chosen in such a way
as to avoid the curvature singularities in the background, and the
quantum instability of metric fluctuations. In addition, we have
included an appropriate radiation backreaction to stabilise the dilaton
in the final, post-Big Bang regime. We have been able to obtain, in this
way, a class of regular cosmological solutions in which the initial
Universe, emerging from the string perturbative vacuum, is first
trapped in a fixed point of high, constant curvature, and then
spontaneously undergoes a smooth transition to the standard FRW
regime. The fixed point is determined by the $\alpha'$ corrections,
while the unstability and the decay of the high curvature regime is
triggered by the quantum loop corrections.

We have determined, in this class of backgrounds, the linearised
equation governing the evolution of tensor metric fluctuations, taking
into account all the $\alpha'$ and loop corrections contributing to the
background regularisation. We have discussed, on this ground, the
amplification of tensor perturbations normalised to the quantum
fluctuations of the vacuum,
and we have estimated their final spectral energy distribution through
analytical and numerical methods.

Our results confirm previous expectations, that the low
frequency modes, crossing the horizon in the low-curvature regime,
are unaffected by higher-order corrections. Also, we have explicitly
checked (through numerical simulations) that the so-called ``string
branch" of the spectrum, associated with the asymptotic, fixed point
regime, can be reliably estimated by means of the low-energy
perturbation equation, that its slope is determined by the
asymptotic constant values of $H$ and $\dot\phi$, and that it is flatter
than at low frequency.

However, when the background regularisation crosses a fixed point,
and when the fixed point is determined by the $\alpha'$ expansion
truncated to first order (like in the class of models of this paper), it
turns out that the slope of the string branch is at least quadratic, and
thus probably too steep to be compatible with future observations by planned
advanced detectors. In the transition regime dominated by the
quantum loops, on the other hand, the influence of the higher-order
corrections on tensor perturbations is more dramatic, and the
corresponding slope may be much flatter than the one of the
preceding string phase.

In summary, the results of this paper agree with the general
expectation that the shape of the spectrum of the relic graviton
background, obtained
in the context of the pre-Big Bang scenario, is strongly
model-dependent. However, it turns out that the formulation of a
complete and consistent regular scenario, associated to a ``visible"
graviton spectrum, may  be more difficult than previously expected
-- at least within a truncated perturbative approach, like the one
adopted in this paper.

\section*{Acknowledgements}
CC is supported by the Swiss NSF, grant No. 83EU-054774 and
ORS/1999041014. EJC was partially supported by PPARC.
 We thank R.~Durrer, M.~Giovannini, S.~Leach,
L.~Mendes, C.~Ungarelli, G.~Veneziano, F.~Vernizzi and D.~Wands
for helpful and stimulating discussions.

\end{document}